\documentclass
[showpacs,nofootinbib,titlepage,12pt,prd,preprint,groupedaddress,noshowkeys]{revtex4}%
\usepackage{amsfonts}
\usepackage{amsmath}
\usepackage{amssymb}
\usepackage{graphicx}%
\begin{document}
\title{Reflection and Transmission at Dimensional Boundaries}
\author{Nelson De Leon and John R. Morris}
\affiliation{Department of Chemistry and Physics, Indiana University Northwest, 3400
Broadway, Gary, Indiana 46408, USA}
\email{ndeleon@iun.edu, jmorris@iun.edu}

\begin{abstract}
An inhomogeneous Kaluza-Klein compactification of a higher dimensional
spacetime may give rise to an effective 4d spacetime with distinct domains
having different sizes of the extra dimensions. The domains are separated by
domain walls generated by the extra dimensional scale factor. The scattering
of electromagnetic and massive particle waves at such boundaries is examined
here for models without warping or branes. We consider the limits
corresponding to thin (thick) domain walls, i.e., limits where wavelengths are
large (small) in comparison to wall thickness. We also obtain numerical
solutions for a wall of arbitrary thickness and extract the reflection and
transmission coefficients as functions of frequency. Results are obtained
which qualitatively resemble those for electroweak domain walls and other
\textquotedblleft ordinary\textquotedblright\ domain walls for 4d theories.

\end{abstract}

\maketitle

\section{Introduction}

The possible existence of unseen extra dimensions could have important
implications for the effective four dimensional physics that we
observe\cite{Kaluza,Klein,ADD,RS1,RS2,DGP,UED1,UED2,Kribs}. Different types of
extra dimensional models include nonorbifolded Kaluza-Klein (KK)
compactifications without branes\cite{Kaluza,Klein}, brane world models with
large\cite{ADD} extra dimensions or warped\cite{RS1,RS2} spacetimes, infinite
extra dimensions\cite{RS2,DGP}, and universal extra dimension
models\cite{UED1,UED2}. Here, we consider KK-type models where a higher
dimensional spacetime without branes or warping is compactified to an
effective 4d spacetime in an \textit{inhomogeneous} way, so that a scale
factor $b(x^{\mu})$ associated with the extra dimension(s) can vary with 4d
position $x^{\mu}$. From a 4d perspective, the scale factor $b$ then appears
as a 4d scalar field with an effective potential $U(b)$\cite{Blau-Guen,CGHW}.
If this effective potential has different minima separated by barriers, the
field $b$ can settle into these different minima at different positions
$x^{\mu}$. The boundaries between the different domains then appear as domain
walls in the 4d theory, where the field $b(x^{\mu})$ varies across the wall,
generally possessing both gradient and potential energy
densities\cite{Blau-Guen,JM1}. The \textquotedblleft gravitational
bags\textquotedblright\ of ref.\cite{DG1} are described by exact analytical
solutions to the field equations of a 6d theory where the extra two dimensions
are compact outside the bag, but become completely decompactified at the
center of the bag ($b\rightarrow\infty$ as $r\rightarrow0$). Dimension
bubbles\cite{JM1,G-M,JM2} are similar nontopological solitons, but filled with
particles and radiation helping to stabilize the bubble, and having slightly
different boundary conditions ($b\rightarrow$ finite as $r\rightarrow0$).
Domain wall networks may give rise to such types of bubbles where the value of
$b$ inside a bubble is different from that outside the
bubble\cite{Blau-Guen,JM1,G-M,JM2}. It is also possible that an evaporating
black hole may spawn a \textquotedblleft modulus bubble\textquotedblright%
\ surrounding the black hole\cite{GSS}. The values of $b$ on different sides
of such a domain wall may both be microscopic, or not. For example, if $b$
takes values $b_{1}$ and $b_{2}$ on different sides of a wall, we may have a
ratio $b_{2}/b_{1}\sim10^{\pm16}$ if $(b_{2}/b_{1})^{\pm1}\sim l_{P}/l_{TeV}$,
i.e., one of the values is characteristic of the Planck scale, $l_{P}\sim
M_{P}^{-1}$, and the other is characteristic of a TeV scale, $l_{TeV}\sim
$TeV$^{-1}$\cite{JM2}. On the other hand, it is possible that the value of $b$
becomes macroscopically large on one side of the wall\cite{Blau-Guen,DG1}.

In the 4d theory that follows from a (brane-free) compactified higher
dimensional theory, the field $b(x^{\mu})$ couples to electromagnetic fields
in the form of a dielectric function $\varepsilon$. Massive particle fields in
the 4d theory have masses which depend on $b$; for example, a particle with a
mass $m_{5}$ in a 5d theory gives rise to a (Kaluza-Klein zero mode) particle
with mass $m=m_{5}/\sqrt{b}$ in the 4d theory\cite{JM2}. Therefore, there will
be a difference in the way that both electromagnetic waves and massive
particle waves will propagate in the two different domains, and we might
anticipate a dependence of the amount of transmission across a dimensional
domain boundary which depends upon the change in the field $b$ and the spatial
rate at which $b$ varies. If, for instance, a massive particle has an energy
$E=(p_{1}^{2}+m_{1}^{2})^{1/2}$ in a domain region where $b=b_{1}$, but across
the wall where $b=b_{2}$ the particle mass is $m_{2}>m_{1},$ then for energies
$E<m_{2}$ the particle will undergo a total reflection, since it is
energetically forbidden in the $b_{2}$ region. For higher energies, $E>m_{2}$,
we expect a partial reflection, with reflection and transmission coefficients
depending upon particle energy\cite{VSbook}. However, we may also anticipate a
dependence upon the domain wall width (as compared to wavelength), as
discovered by Everett\cite{Everett} in his study of wave transmission across
electroweak domain walls. Ayala, Jalilian-Marian, McLerran, and
Vischer\cite{EDW1}, and Farrar and McIntosh\cite{EDW2} have also examined wave
propagation across electroweak domain walls, obtaining energy-dependent
reflection and transmission coefficients. We obtain results for wave
transmission across dimensional boundaries which qualitatively resemble some
of the basic results obtained by refs.\cite{Everett,EDW1,EDW2} for wave
transmission across electroweak domain walls. More specifically, we find that
the reflection coefficient depends upon both $b_{1}$ and $b_{2}$, with
\textit{reflection} being enhanced at \textit{low} energies, where wavelengths
are long in comparison to wall width (thin wall limit). On the other hand,
\textit{transmission} is enhanced and reflection becomes negligible at
\textit{high} energies, where wavelengths are short compared to wall width
(thick wall limit). For the thin wall limit we approximate the wall as a
discontinuous boundary, and in the thick wall limit we consider a slowly
varying field $b(x^{\mu})$. Analytical results can be obtained for these
limiting cases. A numerical study is also made for a smooth, continuous
transition region for a range of energies and wavelengths.

The amount of reflection and transmission of various modes at a dimensional
boundary will generally depend upon the compactification details. Later, as an
example, we show how the qualitative behavior for the reflection of
electromagnetic fields and massive particles differs for the RS1
model\cite{RS1}. Thus, if dimensional boundaries could be probed
experimentally, information about the extra dimensions could be gathered from
the boundary's reflectivity.

In section 2 we present the effective 4d theory that emerges from the KK
reduction of a higher dimensional theory without warping or branes, and
illustrate how the extra dimensional scale factor $b$ appears as a scalar
field in the 4d theory. The reflection and transmission of electromagnetic and
massive particle waves from a thin domain wall is considered in section 3, and
thick walls are treated in section 4. Our numerical study for a domain wall of
arbitrary width is presented in section 5. Section 6 contains a summary and
discussion of results, and expectations concerning scattering from dimensional
boundaries in the visible brane of the RS1 model are mentioned.

\section{The Effective 4d Theory}

We consider a $D=(4+n)$ dimensional spacetime having $n$ compact extra spatial
dimensions. The metric of the $D$ dimensional spacetime is assumed to take a
form%
\begin{equation}
ds_{D}^{2}=\tilde{g}_{MN}dx^{M}dx^{N}=\tilde{g}_{\mu\nu}(x)dx^{\mu}dx^{\nu
}+b^{2}(x^{\mu})\gamma_{mn}(y)dy^{m}dy^{n} \label{e1}%
\end{equation}

where $x^{M}=(x^{\mu},y^{m})$ and $M,N=0,1,2,3,\cdot\cdot\cdot,D-1$ label all
the spacetime coordinates, $\mu,\nu=0,1,2,3$, label the 4d coordinates, and
$m,n$ label those of the compact extra space dimensions. The extra dimensional
scale factor is $b(x^{\mu})$, which is assumed to be independent of the $y$
coordinates and the extra dimensional metric $\gamma_{mn}(y)$ depends upon the
geometry of the extra dimensional space and is related to $\tilde{g}%
_{mn}(x,y)$ by $\tilde{g}_{mn}=b^{2}\gamma_{mn}$.

The action for the $D$ dimensional theory is%
\begin{equation}
S_{D}=\int d^{D}x\sqrt{\left\vert \tilde{g}_{D}\right\vert }\left\{  \frac
{1}{2\kappa_{D}^{2}}\left[  \tilde{R}_{D}[\tilde{g}_{MN}]-2\Lambda\right]
+\mathcal{\tilde{L}}_{D}\right\}  \label{e2}%
\end{equation}

where $\tilde{g}_{D}=\det\tilde{g}_{MN}$, $\tilde{R}_{D}$ is the Ricci scalar
built from $\tilde{g}_{MN}$, $\Lambda$ is a cosmological constant for the $D$
dimensional spacetime, $\mathcal{\tilde{L}}_{D}$ is a Lagrangian for the
fields in the $D$ dimensions, $\kappa_{D}^{2}=8\pi G_{D}=V_{y}\kappa^{2}=8\pi
G$, where $G$ is the 4d gravitational constant, $G_{D}$ is the $D$ dimensional
one, and $V_{y}=\int d^{n}y\sqrt{\left\vert \gamma\right\vert }$ is the
coordinate \textquotedblleft volume\textquotedblright\ of the extra
dimensional space. We use a mostly negative metric signature, $diag(\tilde
{g}_{MN})=(+,-,-,\cdot\cdot\cdot,-)$.

To express the action as an effective 4d action, we borrow the relations used
in ref.\cite{CGHW}:%
\begin{equation}
\sqrt{|\tilde{g}_{D}|}=b^{n}\sqrt{-\tilde{g}}\sqrt{\left\vert \gamma
\right\vert }, \label{e3}%
\end{equation}

\begin{equation}
\tilde{R}[\tilde{g}_{MN}]=\tilde{R}[\tilde{g}_{\mu\nu}]+b^{-2}\tilde{R}%
[\gamma_{mn}]-2nb^{-1}\tilde{g}^{\mu\nu}\tilde{\nabla}_{\mu}\tilde{\nabla
}_{\nu}b-n(n-1)b^{-2}\tilde{g}^{\mu\nu}(\tilde{\nabla}_{\mu}b)(\tilde{\nabla
}_{\nu}b) \label{e4}%
\end{equation}

where the number of extra dimensions $n$ is not to be confused with the tensor
index $n$ and $\tilde{R}[\tilde{g}_{\mu\nu}]$ is the Ricci scalar built from
$\tilde{g}_{\mu\nu}$, etc. The metric $\tilde{g}_{\mu\nu}$ then acts as a 4d
Jordan frame metric. We define $D$ dimensional and 4d gravitation constants by
$2\kappa_{D}^{2}=16\pi G_{D}$ and $2\kappa^{2}=16\pi G$, respectively, which
are related by%
\begin{equation}
\frac{1}{2\kappa^{2}}=\frac{1}{16\pi G}=\frac{V_{y}}{16\pi G_{D}}=\frac{V_{y}%
}{2\kappa_{D}^{2}} \label{e5}%
\end{equation}

Following ref.\cite{CGHW} we consider compact spaces of extra dimensions with
constant curvature and a curvature parameter defined by%
\begin{equation}
k=\frac{\tilde{R}[\gamma_{mn}]}{n(n-1)} \label{e6}%
\end{equation}

Integrating over $y$ in the action of (\ref{e2}), we have\bigskip%
\begin{equation}%
\begin{array}
[c]{ll}%
S & =%
%TCIMACRO{\dint }%
%BeginExpansion
{\displaystyle\int}
%EndExpansion
d^{4}x\sqrt{-\tilde{g}}\left\{  \dfrac{1}{2\kappa^{2}}[b^{n}\tilde{R}%
[\tilde{g}_{\mu\nu}]-2nb^{n-1}\tilde{g}^{\mu\nu}\tilde{\nabla}_{\mu}%
\tilde{\nabla}_{\nu}b-n(n-1)b^{n-2}\tilde{g}^{\mu\nu}(\tilde{\nabla}_{\mu
}b)(\tilde{\nabla}_{\nu}b)\right. \\
& \left.  +n(n-1)kb^{n-2}]+b^{n}\left[  \mathcal{L}_{D}-\dfrac{2\Lambda
}{\kappa^{2}}\right]  \right\}
\end{array}
\label{e7}%
\end{equation}

where $\mathcal{L}_{D}=V_{y}\mathcal{\tilde{L}}_{D}$. We define a 4d Einstein
frame metric $g_{\mu\nu}$ by%
\begin{equation}
\tilde{g}_{\mu\nu}=b^{-n}g_{\mu\nu},\ \ \ \ \ \tilde{g}^{\mu\nu}=b^{n}%
g^{\mu\nu},\ \ \ \ \ \sqrt{-\tilde{g}}=b^{-2n}\sqrt{-g} \label{e8}%
\end{equation}

In terms of the 4d Einstein frame metric the action $S$ in (\ref{e7}) becomes%
\begin{equation}%
\begin{array}
[c]{ll}%
S & =%
%TCIMACRO{\dint }%
%BeginExpansion
{\displaystyle\int}
%EndExpansion
d^{4}x\sqrt{-g}\left\{  \dfrac{1}{2\kappa^{2}}\left[  R[g_{\mu\nu}%
]+\dfrac{n(n+2)}{2}b^{-2}g^{\mu\nu}(\nabla_{\mu}b)(\nabla_{\nu}%
b)+n(n-1)kb^{-(n+2)}\right]  \right. \\
& \ \ \ \ \ \ \ \ \ \ \ \ \ \ \ \ \ \ \ \left.  +b^{-n}\left[  \mathcal{L}%
_{D}-\dfrac{2\Lambda}{\kappa^{2}}\right]  \right\}
\end{array}
\label{e9}%
\end{equation}

where total derivative terms have been dropped. From the $D$ dimensional
source Lagrangian $\mathcal{L}_{D}=V_{y}\mathcal{\tilde{L}}_{D}$ we can define
an effective 4d source Lagrangian $\mathcal{L}_{4}$,%
\begin{equation}
\mathcal{L}_{4}=b^{-n}\mathcal{L}_{D} \label{e10}%
\end{equation}

where, again, $n$ is the number of extra dimensions. Notice that the extra
dimensional scale factor $b(x)$ plays the role of a scalar field in the 4d
theory. It will have a corresponding effective potential $U(b)$ that is
constructed from the curvature and cosmological constant terms in (\ref{e9})
along with terms from $\mathcal{L}_{4}$. When the potential $U$ possesses two
or more minima separated by barriers, domain walls associated with the scalar
field $b(x)$ can appear in the 4d theory. A domain wall interpolates between
two different values of $b_{1}$ and $b_{2}$ on the two sides, and the energy
density and width of the wall are expected to depend on $b_{1}$ and $b_{2}$,
the potential $U(b)$, and how rapidly the field $b$ varies. One can envision a
thin wall where there is a sudden jump between $b_{1}$ and $b_{2}$, a thick
wall where $b(x)$ slowly varies, or intermediate cases where the wall may be
considered as neither thin nor thick.

\subsection{Electromagnetic and Scalar Boson Fields in 4d}

We are interested in the propagation of massive particles and electromagnetic
fields through regions where the size of the extra dimensions, characterized
by the scale factor $b(x)$, changes. We will focus on the simple case of a
free scalar boson as a prototype of a massive particle, and thereby neglect
particle spin. In this case it is easy to see how the 4d particle mass $m$
depends upon the field $b$ for Kaluza-Klein (KK) zero mode bosons. For the
electromagnetic field, it seems natural to adopt a dielectric
approach\cite{G-M}, where the field $b(x)$ gives rise to an effective
permittivity $\varepsilon(x)$. (Attention is restricted to KK zero modes.)

\textit{Electromagnetic Field -- }We write the $D=(4+n)$ dimensional
Lagrangian for the electromagnetic (EM) fields as%
\begin{equation}
\mathcal{L}_{EM}=-\frac{1}{4}\tilde{F}^{\prime MN}\tilde{F}_{MN}^{\prime
}=-\frac{1}{4}\tilde{g}^{MA}\tilde{g}^{NB}F_{AB}^{\prime}F_{MN}^{\prime
},\ \ \ \ \ F_{MN}^{\prime}=\partial_{M}A_{N}^{\prime}-\partial_{N}%
A_{M}^{\prime} \label{e11}%
\end{equation}

We assume the field $b$ to take a value $b_{0}$ in the ambient 4d spacetime,
and then define a rescaled gauge field $A_{M}=b_{0}^{n/2}A_{M}^{\prime}$. For
KK zero modes we assume $A_{M}$ to be independent of $y^{m}$ and set $A_{m}%
=0$. \ In the 4d Einstein frame $\mathcal{L}_{EM}$ becomes $\mathcal{L}%
_{EM}=-\frac{1}{4}b^{2n}F^{\prime\mu\nu}F_{\mu\nu}^{\prime}=-\frac{1}%
{4}(b^{2n}/b_{0}^{n})F^{\mu\nu}F_{\mu\nu}$, with $F_{\mu\nu}=\partial_{\mu
}A_{\nu}-\partial_{\nu}A_{\mu}$. By (\ref{e10}) the effective 4d EM Lagrangian
is%
\begin{equation}
\mathcal{L}_{4,EM}=-\frac{1}{4}\left(  \frac{b}{b_{0}}\right)  ^{n}F^{\mu\nu
}F_{\mu\nu}=-\frac{1}{4}\varepsilon F^{\mu\nu}F_{\mu\nu} \label{e12}%
\end{equation}

where the effective dielectric function is%
\begin{equation}
\varepsilon(x)=\left(  \frac{b(x)}{b_{0}}\right)  ^{n} \label{e13}%
\end{equation}

In ordinary 4d vacuum regions where $b=b_{0}$, we have $\varepsilon=1$.

\textit{Scalar Bosons -- }For the case of a free scalar boson, we start with a
$D$ dimensional Lagrangian%
\begin{equation}
\mathcal{L}_{S}=\tilde{\partial}^{M}\phi^{\ast}\tilde{\partial}_{M}\phi
-V(\phi)=\tilde{g}^{MN}\partial_{M}\phi^{\ast}\partial_{N}\phi-V(\phi)
\label{e14}%
\end{equation}

For KK zero modes $\phi$ is independent of $y^{m}$ and $\mathcal{L}_{S}%
=\tilde{g}^{\mu\nu}\partial_{\mu}\phi^{\ast}\partial_{\nu}\phi-V$. In terms of
the 4d Einstein frame metric from (\ref{e8}) we then have $\mathcal{L}%
_{S}=b^{n}g^{\mu\nu}\partial_{\mu}\phi^{\ast}\partial_{\nu}\phi-V$. From
(\ref{e10}) it follows that the effective 4d Lagrangian is
\begin{equation}
\mathcal{L}_{4,S}=g^{\mu\nu}\partial_{\mu}\phi^{\ast}\partial_{\nu}\phi
-b^{-n}V \label{e15}%
\end{equation}

Therefore a bosonic particle with a mass $\mu_{0}$ in the $4+n$ dimensional
theory appears as a KK zero mode bosonic particle with a 4d mass $m$ given by
$m=b^{-n/2}\mu_{0}$. In terms of the dielectric function $\varepsilon$ and the
mass $m_{0}=b_{0}^{-n/2}\mu_{0}$ we can write
\begin{equation}
m=\varepsilon^{-1/2}m_{0} \label{e16}%
\end{equation}

with $m\rightarrow m_{0}$ as $\varepsilon\rightarrow1$. The 4d mass decreases
in regions of larger $b$ and $\varepsilon$.

\section{Wave Propagation Through Thin Walls}

We consider here the thin wall limit for EM and particle waves, i.e., the
limit in which the effective wavelengths are large in comparison to the wall
width $\delta$, or frequencies are sufficiently small, $\omega\ll1/\delta$.
The cases of EM waves and massive particle waves are considered separately,
but in a similar manner. The transition region where $b(x)$ varies is
idealized as a sharp boundary, i.e., as a planar interface perpendicular to
the $x$-axis.

\subsection{Electromagnetic Waves}

The effects of a rapidly varying $b$ upon EM wave propagation was investigated
in ref.\cite{G-M}. An EM contribution in the form of eq. (\ref{e12}) to the
effective 4d theory can be treated with a dielectric approach where the
dielectric function, or permittivity $\varepsilon$, in a region of space is
given by eq. (\ref{e13}), where $b_{0}$ is the value of the extra dimensional
scale factor in a normal region of 4d spacetime. (The coordinates $y^{m}$
could be rescaled to set $b_{0}=1$, but we leave its value arbitrary.) The
permeability of a region of space is seen from the Maxwell equations to be
$\mu=1/\varepsilon$ so that the index of refraction in a region of space is
$n=\sqrt{\varepsilon\mu}=1$ and the \textquotedblleft
impedance\textquotedblright\ is $Z=\sqrt{\mu/\varepsilon}=1/\varepsilon\propto
b^{-n}$, where $n$ is the number of extra dimensions. In ref.\cite{G-M} it was
found that at the boundary between two different constant values of $b$ given
by $b_{1}$ and $b_{2}$, the reflection ratio for a plane wave of frequency
$\omega$ is given by%
\begin{equation}
A_{ref}=\frac{E_{R}}{E_{I}}=\frac{1-(Z_{T}/Z_{I})}{1+(Z_{T}/Z_{I})}%
=\frac{1-(\varepsilon_{I}/\varepsilon_{T})}{1+(\varepsilon_{I}/\varepsilon
_{T})}=\frac{\varepsilon_{T}-\varepsilon_{I}}{\varepsilon_{T}+\varepsilon_{I}}
\label{e17}%
\end{equation}

where $Z_{I(T)}$ denotes the value of $Z$ in the incident (transmitting)
region, etc. For a reflection coefficient $\mathcal{R}=A_{ref}^{2}$ we have%
\begin{equation}
\mathcal{R}=\left(  \frac{\varepsilon_{T}-\varepsilon_{I}}{\varepsilon
_{T}+\varepsilon_{I}}\right)  ^{2}=\left(  \frac{\varepsilon_{2}%
-\varepsilon_{1}}{\varepsilon_{2}+\varepsilon_{1}}\right)  ^{2}=\left(
\frac{\varepsilon_{2}/\varepsilon_{1}-1}{\varepsilon_{2}/\varepsilon_{1}%
+1}\right)  ^{2} \label{e18}%
\end{equation}

with an invariance under the interchange $\varepsilon_{1}\longleftrightarrow
\varepsilon_{2}$ indicating the same amount of reflection from either side of
the wall. There is a small amount of reflection for $(\Delta\varepsilon
/\varepsilon)^{2}\ll1$, or $\left(  \varepsilon_{2}/\varepsilon_{1}\right)
\approx1$, but a large amount of reflection when either $(\varepsilon
_{2}/\varepsilon_{1})\gg1$ or $(\varepsilon_{2}/\varepsilon_{1})\ll1$.

\subsection{Massive Particles}

We will neglect possible effects due to particle spin and polarization and
therefore examine the reflection and transmission associated with a free
scalar boson field described by eq.(\ref{e15}) with a potential $V=\mu_{0}%
^{2}\phi^{\ast}\phi$. The 4d boson mass $m$ is then given by eq.(\ref{e16}).
We consider the sharp boundary to be located at $x=0$ with $\varepsilon
=\varepsilon_{1}$ for $x<0$ and $\varepsilon=\varepsilon_{2}$ for $x>0$. The
field $\phi$ satisfies the Klein-Gordon equation $\square\phi+m^{2}\phi=0$
with a jump in $m^{2}$ at $x=0$. An incident plane wave of energy $E=\omega$
described by $\phi_{0}$ propagating toward the right is assumed to be incident
from the left ($\varepsilon=\varepsilon_{1}$) on the interface, and a
reflected plane wave $\phi_{1}$ is assumed to propagate back toward the left
in this region. A transmitted wave of energy $\omega$ is transmitted in the
region where $\varepsilon=\varepsilon_{2}$. The plane waveforms are%
\begin{equation}
\phi=\left\{
\begin{array}
[c]{lll}%
\phi_{0}+\phi_{1} & =A_{0}e^{i(p_{1}x-\omega t)}+A_{1}e^{i(-p_{1}x-\omega
t)}, & (x<0)\\
\phi_{2} & =A_{2}e^{i(p_{2}x-\omega t)}, & (x>0)
\end{array}
\right\}  \label{e19}%
\end{equation}

where $\omega^{2}=p_{1}^{2}+m_{1}^{2}=p_{2}^{2}+m_{2}^{2}$ with $p_{1,2}^{2}$
assumed to be nonnegative, and%
\begin{equation}
m^{2}=\left\{
\begin{array}
[c]{cc}%
m_{1}^{2}=\varepsilon_{1}^{-1}m_{0}^{2}, & (x<0)\\
m_{2}^{2}=\varepsilon_{2}^{-1}m_{0}^{2}, & (x>0)
\end{array}
\right\}  \label{e20}%
\end{equation}

Note that if $\omega<m_{2}$, then $p_{2}^{2}$ is negative and the wave is
exponentially attenuated in the region $x>0$. In other words, free particles
are not kinematically allowed into the transmitting region when $\omega<m_{2}%
$, resulting in an effective total reflection.

Requiring continuity of $\phi$ and $\phi^{\prime}=\partial_{x}\phi$ at the
boundary $x=0$ yields%
\begin{equation}
\frac{A_{1}}{A_{0}}=-\frac{(p_{2}-p_{1})}{(p_{2}+p_{1})},\ \ \ \ \ \frac
{A_{2}}{A_{0}}=\frac{2p_{1}}{(p_{2}+p_{1})},\ \ \ \ \ p_{1,2}=(\omega
^{2}-m_{1,2}^{2})^{1/2}\geq0 \label{e21}%
\end{equation}

The $x$-component of the current density is given by $j^{x}=-i\phi^{\ast
}\overleftrightarrow{\partial_{x}}\phi$ so that we can define reflection and
transmission coefficients%
\begin{equation}
\mathcal{R}=-\frac{j_{1}}{j_{0}}=\left(  \frac{A_{1}}{A_{0}}\right)
^{2}=\left(  \frac{p_{2}-p_{1}}{p_{2}+p_{1}}\right)  ^{2}=\left(
\frac{1-(p_{1}/p_{2})}{1+(p_{1}/p_{2})}\right)  ^{2} \label{e22}%
\end{equation}

\begin{equation}
\mathcal{T}=\frac{j_{2}}{j_{0}}=\frac{p_{2}}{p_{1}}\left(  \frac{A_{2}}{A_{0}%
}\right)  ^{2}=\frac{4p_{1}p_{2}}{(p_{1}+p_{2})^{2}}=\frac{4p_{1}}{\left(
1+(p_{1}/p_{2})\right)  ^{2}} \label{e23}%
\end{equation}

with $\mathcal{R+\mathcal{T}}=1$. Again, since $\mathcal{R}$ and $\mathcal{T}$
are symmetric under the interchange of indices, there is the same amount of
reflection and transmission regardless of which side is the incident side.
There is a small amount of reflection when $(\Delta p/p)^{2}\ll1$ or
$p_{2}/p_{1}\approx1$, i.e., when $\varepsilon_{2}/\varepsilon_{1}\approx1$.
On the other hand, there is a large amount of reflection when either
$p_{2}/p_{1}\ll1$ or $p_{2}/p_{1}\gg1$, i.e., when there is a large difference
between $\varepsilon_{1}$ and $\varepsilon_{2}$ with $(\varepsilon
_{2}/\varepsilon_{1})\gg1$ or $(\varepsilon_{2}/\varepsilon_{1})\ll1$. We
therefore get the same qualitative results for reflection/transmission for
both EM waves and particles at a sharp boundary, or for small frequencies or
low energies, $\omega\ll1/\delta$ and $\omega>m_{1,2}$. For the case where the
particle energy $\omega$ is less than the particle mass in the transmitting
region, a free particle is kinematically forbidden in that region, with $\phi$
assumed to be rapidly damped, so that there is effectively total reflection
back into the incident region.

\section{Wave Propagation Through Thick Walls}

In this section we consider the opposite of the thin wall limit, i.e., the
thick wall limit for EM and particle waves where the variation in $b(x)$ in
the transition region is very gradual. This is equivalent to a limit where
wavelengths are small in comparison to the wall thickness, or frequencies and
energies are high, $E=\omega\gg1/\delta$.

\subsection{Electromagnetic Waves}

For sourceless EM fields the effective Lagrangian is given by eqs.(\ref{e12})
and (\ref{e13}), $\mathcal{L}_{4,EM}=-\frac{1}{4}\varepsilon F^{\mu\nu}%
F_{\mu\nu}$ where the effective permittivity is $\varepsilon=(b/b_{0})^{n}$
with the property that $\varepsilon\rightarrow1$ in the normal vacuum. Also,
from the Maxwell field equations we have a permeability $\mu=1/\varepsilon$.
These give rise to an index of refraction $n=\sqrt{\varepsilon\mu}=1$ and
\textquotedblleft impedance\textquotedblright\ $Z=\sqrt{\mu/\varepsilon
}=1/\varepsilon$. Summarizing,%
\begin{equation}
\varepsilon=\frac{1}{\mu}=\left(  \frac{b}{b_{0}}\right)  ^{n},\ \ \ \ n=\sqrt
{\varepsilon\mu}=1,\ \ \ \ Z=\sqrt{\frac{\mu}{\varepsilon}}=\frac
{1}{\varepsilon} \label{e24}%
\end{equation}

We have the inhomogeneous (from $\mathcal{L}_{4,EM}$) and homogeneous (Bianchi
identity) Maxwell equations (source-free case)%
\begin{equation}
\nabla_{\mu}(\varepsilon F^{\mu\nu})=0,\ \ \ \ \ \nabla_{\mu}\ ^{\ast}%
F^{\mu\nu}=0 \label{e25}%
\end{equation}

where the dual tensor is $^{\ast}F_{\mu\nu}=\frac{1}{2}\epsilon_{\mu\nu
\rho\sigma}F^{\rho\sigma}$. With our metric $g_{\mu\nu}=diag(+,-,-,-)$ we have
the field tensors%
\begin{equation}
F_{\mu\nu}=\left(
\begin{array}
[c]{cccc}%
0 & E_{x} & E_{y} & E_{z}\\
-E_{x} & 0 & -B_{z} & B_{y}\\
-E_{y} & B_{z} & 0 & -B_{x}\\
-E_{z} & -B_{y} & B_{x} & 0
\end{array}
\right)  ,\ \ \ \ \ F^{\mu\nu}=\left(
\begin{array}
[c]{cccc}%
0 & -E_{x} & -E_{y} & -E_{z}\\
E_{x} & 0 & -B_{z} & B_{y}\\
E_{y} & B_{z} & 0 & -B_{x}\\
E_{z} & -B_{y} & B_{x} & 0
\end{array}
\right)  \ \ \ \ \ \label{e26}%
\end{equation}

with $E_{x}=E_{1}$, $B_{x}=B_{1}$, etc. being the physical fields and
$F_{ij}=F^{ij}$ and we assume a flat 4d spacetime with $g_{\mu\nu}=\eta
_{\mu\nu}$.

For a simple ansatz we consider waves propagating in the $+x$ direction and we
take $\varepsilon=\varepsilon(x)$ to be a smooth, slowly varying, although
somewhat arbitrary, function. We take the field components $E_{y}(x,t)$ and
$B_{z}(x,t)$ to be nonvanishing and complex, in general, with $\vec{H}=\vec
{B}/\mu=\varepsilon\vec{B}$. There is then a nonvanishing Poynting vector%
\begin{equation}
\vec{S}=\operatorname{Re}\left(  \vec{E}\times\vec{H}^{\ast}\right)
=\varepsilon\operatorname{Re}\left(  \vec{E}\times\vec{B}^{\ast}\right)
\label{e27}%
\end{equation}

with a nonvanishing component $S_{x}$. The field eqs.(\ref{e25}) can then be
written in the form%
\begin{equation}
\dot{E}+B^{\prime}+\gamma^{\prime}B=0,\ \ \ \ \ E^{\prime}=-\dot{B}
\label{e28}%
\end{equation}

where $E=E_{y}$, $B=B_{z}$, an overdot represents differentiation with respect
to $t$, a prime stands for differentiation with respect to $x$, and we have
defined $\gamma=\ln\varepsilon$. These equations can be used to obtain the
wave equation%
\begin{equation}
B^{\prime\prime}-\ddot{B}+\gamma^{\prime\prime}B+\gamma^{\prime}B^{\prime}=0
\label{e29}%
\end{equation}

\textit{Approximate solutions} -- We consider the special case where the scale
factor $b(x)$, $\varepsilon(x)$ and $\gamma(x)$ are very slowly varying
functions of $x$ in a planar domain wall oriented perpendicular to the $x$
axis. The overall change in $\varepsilon$ can be arbitrarily large, but the
rate of change must be sufficiently small for the frequencies under
consideration. The magnetic field is assumed to be of the form%
\begin{equation}
B(x,t)=Ae^{i\phi(x)}e^{-i\omega t} \label{e30}%
\end{equation}

where the amplitude $A$ is a real constant. The wave equation (\ref{e29}) then
gives an equation for the phase function $\phi$,%
\begin{equation}
i\phi^{\prime\prime}-\phi^{\prime2}+\omega^{2}+\gamma^{\prime\prime}%
+i\phi^{\prime}\gamma^{\prime}=0 \label{e31}%
\end{equation}

Now, for the ordinary case where $\varepsilon=const$, or $\gamma^{\prime}=0$,
the wave eq. for $B(x,t)$ reduces to the ordinary wave equation,
$B^{\prime\prime}-\ddot{B}=0$, giving $\phi(x)=\pm\omega x$, $\phi^{\prime
}=\pm\omega$, $\phi^{\prime\prime}=0$. Therefore, for the case of a very
slowly changing $\gamma$ and $\phi^{\prime}$, we, as a first approximation,
drop the terms involving $\gamma^{\prime\prime}$ and $\phi^{\prime\prime}$ and
approximate $\phi^{\prime}\gamma^{\prime}\approx\pm\omega\gamma^{\prime}$ in
(\ref{e31}), so that it reduces to the approximate equation%
\begin{equation}
(\phi_{\pm}^{\prime})^{2}=\omega^{2}\pm i\omega\gamma^{\prime},\ \ \ \ \ \phi
_{\pm}^{\prime}=\pm\omega\sqrt{1\pm i\frac{\gamma^{\prime}}{\omega}}
\label{32}%
\end{equation}

i.e., we use $\phi^{\prime}\gamma^{\prime}\approx+\omega\gamma^{\prime}$ for
the $\phi_{+}$ solution and use $\phi^{\prime}\gamma^{\prime}\approx
-\omega\gamma^{\prime}$ for the $\phi_{-}$ solution, giving $\phi_{+}^{\prime
}=\omega\sqrt{1+i\dfrac{\gamma^{\prime}}{\omega}}$ and $\phi_{-}^{\prime
}=-\omega\sqrt{1-i\dfrac{\gamma^{\prime}}{\omega}}$ for the two solutions. We
consider frequencies for which $|\gamma^{\prime}/\omega|\ll1$ so that the
approximate solution is given by%
\begin{equation}
\phi_{\pm}^{\prime}=\pm\omega+i\frac{\gamma^{\prime}}{2},\ \ \ \ \ (|\gamma
^{\prime}|\ll\omega) \label{e33}%
\end{equation}

Integrating this gives%
\begin{equation}
\phi_{\pm}(x)-\phi_{\pm}(x_{0})=\pm\omega(x-x_{0})+\frac{i}{2}[\gamma
(x)-\gamma(x_{0})] \label{e34}%
\end{equation}

We choose to set the constants $\phi_{\pm}(x_{0})=\pm\omega x_{0}$ so that the
solution simplifies to%
\begin{equation}
\phi_{\pm}(x)=\pm\omega x+\frac{i}{2}\ln\left(  \frac{\varepsilon
(x)}{\varepsilon(x_{0})}\right)  =\pm\omega x+i\ln\left(  \frac{\varepsilon
}{\varepsilon_{0}}\right)  ^{1/2} \label{e35}%
\end{equation}

where $\varepsilon=\varepsilon(x)$ and $\varepsilon_{0}=\varepsilon(x_{0})$.
We then have $e^{i\phi_{\pm}}=\left(  \frac{\varepsilon}{\varepsilon_{0}%
}\right)  ^{-1/2}e^{\pm i\omega x}$ and%
\begin{equation}
B_{\pm}(x,t)=A\left(  \frac{\varepsilon}{\varepsilon_{0}}\right)
^{-1/2}e^{\pm i\omega x}e^{-i\omega t} \label{e36}%
\end{equation}

This describes waves propagating in the $\pm x$ directions traveling at the
speed of light in vacuum ($\omega/k=1$) with an effective amplitude $A\left(
\frac{\varepsilon}{\varepsilon_{0}}\right)  ^{-1/2}$ which varies with $x$.

From the field eqs. (\ref{e28}) we have $\dot{E}+B^{\prime}+\gamma^{\prime
}B=0$ so that, using $\dot{E}=-i\omega E$ and $B_{\pm}^{\prime}=i\phi_{\pm
}^{\prime}B_{\pm}$ we obtain the electric field%
\begin{equation}%
\begin{array}
[c]{ll}%
E_{\pm} & =-\dfrac{i}{\omega}(B_{\pm}^{\prime}+\gamma^{\prime}B_{\pm})=\left(
\pm1-i\dfrac{\gamma^{\prime}}{2\omega}\right)  B_{\pm}%
\end{array}
\label{e37}%
\end{equation}

The approximate solutions for the EM fields are then%
\begin{equation}%
\begin{array}
[c]{ll}%
B_{\pm}(x,t)=B_{z,\pm} & =A\left(  \dfrac{\varepsilon}{\varepsilon_{0}%
}\right)  ^{-1/2}e^{\pm i\omega x}e^{-i\omega t}\\
E_{\pm}(x,t)=E_{y,\pm} & =\left(  \pm1-i\dfrac{\gamma^{\prime}}{2\omega
}\right)  A\left(  \dfrac{\varepsilon}{\varepsilon_{0}}\right)  ^{-1/2}e^{\pm
i\omega x}e^{-i\omega t}%
\end{array}
\label{e38}%
\end{equation}

\textit{Reflection and Transmission} -- The EM energy flow is indicated by the
Poynting vector $\vec{S}=\operatorname{Re}(\vec{E}\times\vec{H}^{\ast})$, with
$\vec{H}=\vec{B}/\mu=\varepsilon\vec{B}$, so that%
\begin{equation}
S_{x}=\varepsilon\operatorname{Re}(\vec{E}\times\vec{B}^{\ast})_{x}%
=\frac{\varepsilon}{2}(E_{y}B_{z}^{\ast}+E_{y}^{\ast}B_{z}) \label{e39}%
\end{equation}

with $E_{\pm}=(\pm1-i\frac{\gamma^{\prime}}{2\omega})B_{\pm}$. We have
$E_{+}B_{+}^{\ast}=(1-i\frac{\gamma^{\prime}}{2\omega})B_{+}B_{+}^{\ast}$,
$E_{-}B_{-}^{\ast}=(-1-i\frac{\gamma^{\prime}}{2\omega})B_{-}B_{-}^{\ast}$,
etc., so that%
\begin{equation}
E_{+}B_{+}^{\ast}+E_{+}^{\ast}B_{+}=2B_{+}^{\ast}B_{+},\ \ \ \ E_{-}%
B_{-}^{\ast}+E_{-}^{\ast}B_{-}=-2B_{-}^{\ast}B_{-} \label{e40}%
\end{equation}

giving%
\begin{equation}
(S_{x})_{\pm}=\varepsilon\operatorname{Re}(\vec{E}_{\pm}\times\vec{B}_{\pm
}^{\ast})=\pm\varepsilon\left\vert \vec{B}_{\pm}\right\vert ^{2}%
=\pm\varepsilon\left[  A^{2}\frac{\varepsilon_{0}}{\varepsilon}\right]
=\pm\varepsilon_{0}A^{2} \label{e41}%
\end{equation}

The Poynting vector $S_{x}\propto A^{2}$ is therefore $x$ independent,
indicating that no energy is lost by reflection or absorption by the traveling
wave, i.e., within the approximation we have used, the transmission amplitude
is unity and the reflection amplitude is zero, giving transmission and
reflection coefficients $\mathcal{T}\approx1$, $\mathcal{R}\approx0$ for high
frequencies $\omega\gg\left\vert \gamma^{\prime}\right\vert $. For a linear
approximation of $\gamma(x)$ in a domain wall of width $\delta$, we have
$\gamma^{\prime}\approx(\gamma-\gamma_{0})/(x-x_{0})=\ln(\varepsilon
/\varepsilon_{0})/\delta$ and the condition $\omega\gg|\gamma^{\prime}|$
translates into $\omega\gg|\frac{\ln(\varepsilon/\varepsilon_{0})}{\delta}|$.
So, for sufficiently high frequencies the wall is transparent,%
\begin{equation}
\mathcal{T}\approx1,\ \ \mathcal{R}\approx0,\ \ \ \ \ \omega\gg\left\vert
\gamma^{\prime}\right\vert \sim\left\vert \frac{\ln(\varepsilon/\varepsilon
_{0})}{\delta}\right\vert \label{e42}%
\end{equation}

This resembles the situation found for \textquotedblleft
ordinary\textquotedblright\ electroweak domain walls separating different
electroweak phases\cite{Everett,EDW1,EDW2}.

\subsection{Massive Particles}

We now focus again on \textquotedblleft free\textquotedblright\ scalar bosons
obeying the Klein-Gordon equation (KGE) $\square\phi+m^{2}(x)\phi=0$, or%
\begin{equation}
\ddot{\phi}-\phi^{\prime\prime}+m^{2}(x)\phi=0 \label{e43}%
\end{equation}
where $m^{2}(x)=b^{-n}(x)\mu_{0}^{2}=\varepsilon^{-1}(x)m_{0}^{2}$ is a very
slowly varying function of $x$. We write the scalar field as%
\begin{equation}
\phi(x,t)=Ae^{i\psi(x)}e^{-i\omega t} \label{e44}%
\end{equation}

where the amplitude $A$ is a real constant, $\psi(x)$ is a phase function to
be determined, and $\omega=E=\sqrt{p^{2}+m^{2}}$ is a fixed, constant energy
(although the momentum $p$ and mass $m$ vary with $x$, in general). For the
case that $m=const$, the solutions are $\phi=Ae^{\pm ipx}e^{-i\omega t}$. For
the more general case, using eq.(\ref{e43}), the KGE gives an equation for the
phase function,%
\begin{equation}
i\psi^{\prime\prime}-\psi^{\prime2}+(\omega^{2}-m^{2})=0 \label{e45}%
\end{equation}

Again, for the case that $m=const$, we have $\psi=\pm px$ with $\psi
^{\prime\prime}=0$.

\textit{Approximate solutions} -- For slowly varying $m^{2}(x)$ let us assume
$\left\vert \psi^{\prime\prime}\right\vert \ll\psi^{\prime2}$ so that we can
drop the $\psi^{\prime\prime}$ term in eq.(\ref{e45}). This leaves us with
$\psi^{\prime}=\pm\omega\left(  1-m^{2}/\omega^{2}\right)  ^{1/2}$ and we find
that for a sufficiently mildly varying function $m(x)=m_{0}/\sqrt
{\varepsilon(x)}\propto b^{-n/2}(x)$ or at a sufficiently high energy $\omega$
our assumption $\left\vert \psi^{\prime\prime}\right\vert \ll\psi^{\prime2}$
is valid, i.e.,
\begin{equation}
\left\vert \frac{\psi^{\prime\prime}}{\psi^{\prime2}}\right\vert =\left\vert
\frac{mm^{\prime}}{\omega^{3}}\right\vert \frac{1}{\left(  1-\frac{m^{2}%
}{\omega^{2}}\right)  ^{3/2}}\ll1 \label{e46}%
\end{equation}

Since $\omega\geq m$, this condition will be satisfied when the first factor
on the right hand side is small, i.e., $\left\vert mm^{\prime}/\omega
^{3}\right\vert =\left\vert \frac{nb^{\prime}}{2b}\frac{m^{2}}{\omega^{3}%
}\right\vert \lesssim\left\vert \frac{b^{\prime}/b}{\omega}\right\vert \ll1$.
This is satisfied for an arbitrary function $m(x)$ at a sufficiently high
energy $\omega$, where $\left\vert m^{\prime}/m\right\vert \ll\omega$ or
$\left\vert b^{\prime}/b\right\vert \ll\omega$. We have $\psi^{\prime2}%
\approx(\omega^{2}-m^{2})$ and can then write, approximately,%
\begin{equation}
\psi_{\pm}(x)=\pm\left\{  \omega x-\frac{1}{2}\int_{x_{0}}^{x}\frac{m^{2}%
}{\omega}dx\right\}  \label{e47}%
\end{equation}

and%
\begin{equation}
\phi_{\pm}(x,t)=A_{\pm}e^{i\psi_{\pm}(x)}e^{-i\omega t} \label{e48}%
\end{equation}

\textit{Reflection and Transmission} -- The current density $j^{\mu}%
=i\phi^{\ast}\overleftrightarrow{\partial^{\mu}}\phi$ then gives,
approximately,%
\begin{equation}
j_{\pm}^{x}=2A^{2}\psi_{\pm}^{\prime}=\pm2A_{\pm}^{2}(\omega^{2}-m^{2}%
)^{1/2}=\pm2A_{\pm}^{2}\omega\left(  1-\frac{m^{2}}{\omega^{2}}\right)  ^{1/2}
\label{e49}%
\end{equation}

Denoting $m(-\infty)=m_{1}$ and $m(+\infty)=m_{2}$, we write the transmission
coefficient as%
\begin{equation}
\mathcal{\mathcal{T}}=\frac{j_{2+}^{x}}{j_{1+}^{x}}=\frac{j_{+}^{x}(\infty
)}{j_{+}^{x}(-\infty)}=\left(  \frac{\omega^{2}-m_{2}^{2}}{\omega^{2}%
-m_{1}^{2}}\right)  ^{1/2}\approx\left[  1-\frac{(m_{2}^{2}-m_{1}^{2})}%
{\omega^{2}}\right]  ,\ \ \ \ m^{2}/\omega^{2}\ll1 \label{e50}%
\end{equation}

Therefore, up to small corrections of $O(m^{2}/\omega^{2})$, we have a
transmission coefficient of unity;%
\begin{equation}
\mathcal{R}\approx0,\ \ \ \mathcal{T}\approx1,\ \ \ \ \ \omega\gg
m,\ \ \ \omega\gg\left\vert m^{\prime}/m\right\vert \label{e51}%
\end{equation}

From (\ref{e42}) and (\ref{e51}) we conclude that thick walls are essentially
transparent to particle and EM radiation at very high energies.

\section{Numerical Results}

We have made a numerical study of the reflection and transmission of
electromagnetic and matter waves at a dimensional boundary characterized by a
smooth, continuous transition region with a dielectric function $\varepsilon
(x)$ which tends to a value $\varepsilon\rightarrow\varepsilon_{1}$ as
$x\rightarrow-\infty$ and a value $\varepsilon\rightarrow\varepsilon_{2}$ as
$x\rightarrow+\infty$. We examined monochromatic waves that solve the field
equations and have used various $\varepsilon_{1}$ and $\varepsilon_{2}$
values. From the currents for the waves, the reflection coefficient
$\mathcal{R}$ and transmission coefficient $\mathcal{T}$ were computed. These
coefficients can then be displayed as functions of $\omega$ for various values
of $\varepsilon_{1}$ and/or $\varepsilon_{2}$. The numerical procedure that
the code is based upon is described in the appendix.

For a simple representation of a smooth transition region interpolating
between $\varepsilon_{1}$ at $x=-\infty$ and $\varepsilon_{2}$ at $x=+\infty$,
we take $\varepsilon(x)$ to be given by the function%
\begin{equation}
\varepsilon(x)=C+D\tanh\left(  \frac{x}{\delta}\right)  =\frac{1}{2}\left\{
(\varepsilon_{1}+\varepsilon_{2})+(\varepsilon_{2}-\varepsilon_{1}%
)\tanh\left(  \frac{x}{\delta}\right)  \right\}  \label{e52}%
\end{equation}

where $C=(\varepsilon_{1}+\varepsilon_{2})/2$ and $D=(\varepsilon
_{2}-\varepsilon_{1})/2$. We also define $\gamma(x)=\ln\varepsilon(x)$, as
before. The parameter $\delta$ characterizes the width of the domain wall
function $\varepsilon(x)$. Dimensioned quantities are rescaled by factors of
$\delta$ to give dimensionless ones. For example, we have dimensionless
quantities (denoted with an overbar)%
\begin{equation}
\bar{x}=\frac{x}{\delta},\ \ \bar{\omega}=\omega\delta,\ \ \bar{m}%
=m\delta,\ \ \bar{k}=k\delta,\ \ \bar{\lambda}=\frac{\lambda}{\delta}%
=\frac{2\pi}{\bar{k}} \label{e53}%
\end{equation}

where $k=\sqrt{\omega^{2}-m^{2}(x)}$ is the particle momentum and $\omega=E$
is the particle energy. (The equations with dimensionless parameters can be
obtained from those with dimensioned parameters by simply setting $\delta=1$
and regarding the dimensioned parameters as dimensionless ones.)

\subsection{Electromagnetic Waves}

We write the electric and magnetic fields, respectively, as
$E(x,t)=E(x)e^{-i\omega t}$ and $B(x,t)=B(x)e^{-i\omega t}$, with
$E(x)=E_{y}(x)$ and $B(x)=B_{z}(x)$ being complex-valued, in general. The
field equations are given by (\ref{e28}) and lead to the wave equation
(\ref{e29}) for $B(x,t)$. The monochromatic ansatz leads to the wave equation
for $B(x)$:%
\begin{equation}
B^{\prime\prime}+\omega^{2}B+\gamma^{\prime\prime}B+\gamma^{\prime}B^{\prime
}=0 \label{e54}%
\end{equation}

where again the prime stands for differentiation with respect to $x$. Using
eq.(\ref{e28}) $E(x)$ is given by $E=-\frac{i}{\omega}(B^{\prime}%
+\gamma^{\prime}B)$. From eq.(\ref{e39}) the $x$ component of the Poynting
vector is%
\begin{equation}
S=\varepsilon\operatorname{Re}(EB^{\ast})=\frac{\varepsilon}{2}(EB^{\ast
}+E^{\ast}B) \label{e55}%
\end{equation}

The wave equation (\ref{e54}) is to be solved numerically, subject to boundary
conditions, which are posted in the form of solutions to the wave equation in
the asymptotic regions. That is, as $x\rightarrow\pm\infty$, we have
$\gamma^{\prime}\rightarrow0$ and solutions for $B$ and $E$ are left-moving
and right-moving plane wave solutions:%
\begin{equation}
B(x)=\left\{
\begin{array}
[c]{ll}%
e^{i\omega x}+A_{1}e^{i\delta_{1}}e^{-i\omega x}, & x\rightarrow-\infty\\
A_{2}e^{i\delta_{2}}e^{i\omega x}, & x\rightarrow+\infty
\end{array}
\right\}  \label{e56}%
\end{equation}

and%
\begin{equation}
E(x)=\left\{
\begin{array}
[c]{ll}%
e^{i\omega x}-A_{1}e^{i\delta_{1}}e^{-i\omega x}, & x\rightarrow-\infty\\
A_{2}e^{i\delta_{2}}e^{i\omega x}, & x\rightarrow+\infty
\end{array}
\right\}  \label{e57}%
\end{equation}

where $A_{1}$ and $\delta_{1}$ are the (real) amplitude and phase constant,
respectively, for the reflected wave, and $A_{2}$ and $\delta_{2}$ are those
for the transmitted wave and the incident fields are $E_{inc}=B_{inc}%
=e^{i\omega x}$. The EM energy momentum is conserved with $S^{\prime}=0$. From
eqs.(\ref{e55}) -- (\ref{e57}) we have%
\begin{equation}
S=\left\{
\begin{array}
[c]{ll}%
S_{inc}+S_{-}(-\infty)=\varepsilon_{1}(1-A_{1}^{2}), & x\rightarrow-\infty\\
S_{+}(+\infty)=\varepsilon_{2}A_{2}^{2}, & x\rightarrow+\infty
\end{array}
\right\}  \label{e58}%
\end{equation}

where $S_{inc}=\varepsilon_{1}$ is the energy-momentum of the incident beam,
$S_{-}=S-S_{inc}$ and $S_{+}$ represent the reflected and transmitted
energy-momentum flow, respectively. The solutions should respect the condition
$S^{\prime}=0$, or%
\begin{equation}
\varepsilon_{1}(1-A_{1}^{2})=\varepsilon_{2}A_{2}^{2} \label{e59}%
\end{equation}

The reflection and transmission coefficients are then defined by%
\begin{equation}
\mathcal{R=}-\frac{S_{-}(-\infty)}{S_{inc}}=A_{1}^{2},\ \ \ \ \ \mathcal{T}%
=\frac{S_{+}(+\infty)}{S_{inc}}=\frac{\varepsilon_{2}}{\varepsilon_{1}}%
A_{2}^{2} \label{e60}%
\end{equation}

The condition given by eq.(\ref{e59}) then implies that $\mathcal{R+T}=1$.

Results obtained for the reflection and transmission coefficients, as
functions of dimensionless frequency $\bar{\omega}=\omega\delta$, are
illustrated in figure 1. These coefficients are monotonic functions and have
limiting values that are in agreement with the analytical results obtained
earlier for thin walls ($\omega\rightarrow0$, $\lambda\gg\delta$), given by
eq.(\ref{e18}), and for thick walls ($\omega\gg\delta^{-1}$, $\lambda\ll
\delta$), given by eq.(\ref{e42}). From figure 1 the reflection is seen to be
substantially reduced below its maximum value around a frequency $\omega
\sim\delta^{-1}$, i.e., $\bar{\omega}\sim1$. The numerical study indicates
that the functions $\mathcal{R}$ and $\mathcal{T}$ are invariant under the
interchange $\varepsilon_{1}\leftrightarrow\varepsilon_{2}$, so that the
amount of reflection does not depend on whether the beam is incident from the
left or the right.

\subsection{Massive Particles}

The boson field obeys the Klein-Gordon equation $\square\phi+m^{2}(x)\phi=0$
and we take $\phi(x,t)=\phi(x)e^{-i\omega t}$. From eq.(\ref{e16}) we have
$m^{2}(x)=m_{0}^{2}/\varepsilon(x)$, where $m_{0}$ is a constant. We define a
dimensionless mass parameter $\bar{m}_{0}=m_{0}\delta$ and set $\bar{m}_{0}=1$
for convenience, letting $\delta^{-1}$ set a mass scale. The Klein-Gordon
equation for $\phi(x)$ can then be written in terms of dimensionless
parameters, and takes the form%
\begin{equation}
\phi^{\prime\prime}-\left(  \bar{\omega}^{2}-\frac{1}{\varepsilon(x)}\right)
\phi=0 \label{e61}%
\end{equation}

or, $\phi^{\prime\prime}(x)-\bar{k}^{2}(x)\phi(x)=0$, where $\bar{m}_{0}=1$
and
\begin{equation}
\bar{k}=\sqrt{\bar{\omega}^{2}-\bar{m}^{2}(x)}=\sqrt{\bar{\omega}^{2}-\frac
{1}{\varepsilon(x)}} \label{e62}%
\end{equation}

In the asymptotic regions we have $\varepsilon\rightarrow\varepsilon_{1}$ as
$x\rightarrow-\infty$ and $\varepsilon\rightarrow\varepsilon_{2}$ as
$x\rightarrow+\infty$, so that%
\begin{equation}
\bar{k}=\left\{
\begin{array}
[c]{cc}%
\bar{k}_{1}=\sqrt{\bar{\omega}^{2}-\dfrac{1}{\varepsilon_{1}}}, &
x\rightarrow-\infty\\
\bar{k}_{2}=\sqrt{\bar{\omega}^{2}-\dfrac{1}{\varepsilon_{2}}}, &
x\rightarrow+\infty
\end{array}
\right\}  \label{e63}%
\end{equation}

and%
\begin{equation}
\phi(x)=\left\{
\begin{array}
[c]{ll}%
\phi_{inc}+\phi_{1}=e^{ik_{1}x}+A_{1}e^{i\delta_{1}}e^{-ik_{1}x}, &
x\rightarrow-\infty\\
\phi_{2}=A_{2}e^{i\delta_{2}}e^{ik_{2}x}, & x\rightarrow+\infty
\end{array}
\right\}  \label{e64}%
\end{equation}

The bosonic current density $j^{\mu}(x,t)=i\phi^{\ast}\overleftrightarrow
{\partial^{\mu}}\phi$ is conserved, so that $j^{\prime}(x)=0$, where the $x$
component of the current is%
\begin{equation}
j(x)=-i(\phi^{\ast}\phi^{\prime}-\phi^{\ast\prime}\phi)=2\operatorname{Im}%
(\phi^{\ast}\phi^{\prime}) \label{e65}%
\end{equation}

From eqs.(\ref{e64}) and (\ref{e65}) the asymptotic currents are%
\begin{equation}
j=\left\{
\begin{array}
[c]{ll}%
j(-\infty)=j_{inc}+j_{-}(-\infty)=2k_{1}(1-A_{1}^{2}), & x\rightarrow-\infty\\
j(+\infty)=j_{+}(+\infty)=2k_{2}A_{2}^{2}, & x\rightarrow+\infty
\end{array}
\right\}  \label{e66}%
\end{equation}

where $j_{-}=-2k_{1}A_{1}^{2}$ is the current of the reflected wave, $j_{+}$
is the current of the transmitted right-moving wave, and $j_{inc}=2k_{1}$ is
the current of the incident beam. We define the reflection and transmission
coefficients as%
\begin{equation}
\mathcal{R}=\frac{-j_{-}(-\infty)}{j_{inc}}=A_{1}^{2},\ \ \ \ \ \mathcal{T}%
=\frac{j_{+}(+\infty)}{j_{inc}}=\frac{k_{2}}{k_{1}}A_{2}^{2} \label{e67}%
\end{equation}

From $j^{\prime}=0$ it follows that the solutions must satisfy the condition
$\mathcal{R}+\mathcal{T}=1$.

The results for the reflection and transmission coefficients as functions of
frequency $\bar{\omega}=\omega\delta$ obtained from the numerical solutions
are shown in figure 2. Again, these are monotonic functions, as expected, and
$\mathcal{T}\rightarrow1$ in the high energy limit where the thick wall
approximation becomes valid (see eq.(\ref{e51})) for $\lambda\ll\delta$, or
$\bar{\omega}\gg1/\varepsilon$. We have chosen $\varepsilon_{1}=1$ in figure
2, so that $k\geq0$ implies that $\bar{\omega}\geq1$. For $\bar{\omega}$ only
a few percent larger than unity, there is an apparent deviation from the
analytical results obtained in eqs.(\ref{e22}) and (\ref{e23}) for the thin
wall approximation where we used a sharp boundary. This is understood by
writing the thin wall approximation $\lambda\gg\delta$ in terms of
$\bar{\omega}$, i.e., $\dfrac{1}{\sqrt{\varepsilon}}\leq\bar{\omega}\ll
2\pi\sqrt{1+\dfrac{1}{4\pi^{2}\varepsilon}}$. This inequality is hard to
satisfy because of the presence of the lower bound, $\bar{\omega}\geq1$ in the
regions where a propagating wave is kinematically allowed, making the range of
$\bar{\omega}$ rather restricted for the case where $\varepsilon\geq1$.
However, in the limit that $\bar{\omega}\rightarrow1$, ($k_{1}\rightarrow0$)
we numerically determine that $\mathcal{R}\rightarrow1$, $\mathcal{T}%
\rightarrow0$ as implied by eqs.(\ref{e22}) and (\ref{e23}).Therefore, the
analytical thin wall results are approached in the limit $\bar{\omega
}\rightarrow1$, but quickly deviate somewhat from the thin wall approximation
in the plots of figure 2. As with the case of electromagnetic waves, the
numerical study indicates that the functions $\mathcal{R}$ and $\mathcal{T}$
are invariant under the interchange $\varepsilon_{1}\leftrightarrow
\varepsilon_{2}$, so that the amount of reflection is independent of whether
the beam is incident from the left or the right.

\section{Summary and Discussion}

We have considered a situation wherein an \textit{inhomogeneous}
compactification of a $4+n$ dimensional spacetime, without warping or branes,
with $n$ compact extra space dimensions, gives rise to an effective 4d
spacetime with distinct domains having different sizes of the extra
dimensions. From a 4d point of view these domains are separated by domain
walls arising from a 4d scalar field $b(x^{\mu})$, which is also the scale
factor for the extra dimensions in the higher dimensional spacetime. The field
$b$ giving rise to a domain wall takes an asymptotic value of $b_{1}$ on one
side of the wall and an asymptotic value of $b_{2}$ on the other side, so that
the domain wall serves as a dimensional boundary. We have focused on the
reflection and transmission of both electromagnetic waves and massive bosonic
particle waves across such dimensional boundaries. This has been done by
examining the limiting cases of thin (thick) walls, i.e., wall thicknesses
that are small (large) in comparison to the wavelengths of the propagating
waves. A convenient parameter for describing the sizes of the extra dimensions
is the \textquotedblleft dielectric function\textquotedblright\ $\varepsilon
(x)=\left(  b(x)/b_{0}\right)  ^{n}$ where $b$ takes an asymptotic, constant
value of $b_{0}$ in a region of ordinary 4d vacuum.

The results we obtain for the reflection and transmission across dimensional
boundaries is qualitatively similar to those obtained for ordinary domain
walls in a 4d theory. Specifically, we find that at very high energies the
boundaries are essentially transparent to EM and particle radiation, while at
low energies, the degree of reflectivity can be quite high if either
$\varepsilon_{2}/\varepsilon_{1}\gg1$ or $\varepsilon_{2}/\varepsilon_{1}\ll
1$, that is, if there is a dramatic change in the size of the extra dimensions
across the boundary. This could be realized, for example, when the extra
dimensions remain microscopically small in both regions while $\left(
b_{2}/b_{1}\right)  ^{\pm1}\sim l_{P}/l_{TeV}$, where $l_{p}\sim M_{P}^{-1}$
is the Planck length and $l_{TeV}\sim TeV^{-1}$. For particles with nonzero
masses there is a threshold energy ($\omega\geq\omega_{\min}$, for which
$k_{1}$ or $k_{2}$ becomes zero) for propagating waves, with $\mathcal{R}%
\rightarrow1$ as $\omega\rightarrow\omega_{\min}$. Our numerical study
substantiates these results for the case of a domain wall of arbitrary width
$\delta$, with $\mathcal{R}$ and $\mathcal{T}$ being monotonic functions
smoothly connecting the thin wall and thick wall approximations for various wavelengths.

We note, however, that the results obtained here are valid only for extra
dimensional models having compact extra dimensions without warping or branes.
Braneworld models can exhibit different qualitative behaviors for the
reflection and transmission of massless or massive modes. As an example,
consider the RS1 model\cite{RS1} consisting of one extra dimension
compactified on an $S^{1}/Z_{2}$ orbifold. The background spacetime metric is%
\begin{equation}
ds^{2}=e^{-2kr|\phi|}g_{\mu\nu}dx^{\mu}dx^{\nu}-r^{2}d\phi^{2} \label{e68}%
\end{equation}

where $-\pi\leq\phi\leq\pi$ with $(x^{\mu},\phi)$ and $(x^{\mu},-\phi)$
identified. The two 3-branes are located at $\phi=0$ (hidden brane) and
$\phi=\pi$ (visible brane). The parameter $k$ is a constant, $e^{-2kr|\phi|}$
is the warp factor, and $r$ is the radius of the compactified extra dimension.
Let us consider a situation wherein the radius $r$ becomes a function of
$x^{\mu}$, corresponding to a case of inhomogeneous compactification. If the
spatial variation of $r$ is mild, we would expect the basic results obtained
from the RS1 model with constant $r$ to be approximately valid, at least
qualitatively. In particular, a physical particle mass $m$ on the visible
brane is related to the mass parameter $m_{0}$ appearing in the 5d theory by%
\begin{equation}
m=e^{-kr\pi}m_{0} \label{e69}%
\end{equation}

Therefore, if $r$ varies with position $x$, the mass $m$ on the visible brane
is smaller for larger $r$, i.e., for a larger extra dimension. This is the
same type of basic behavior found above for our unwarped, brane-free models,
and we therefore expect the same qualitative type of reflection behavior for
massive particles at a dimensional boundary. The story is different, however,
for electromagnetic fields. To see this, we write the contribution to the EM
fields on the visible brane as%
\begin{equation}
S_{v}=\int d^{4}x\sqrt{-g_{v}}\left(  -\frac{1}{4}g_{v}^{\mu\alpha}g_{v}%
^{\nu\beta}F_{\alpha\beta}F_{\mu\nu}\right)  \label{e70}%
\end{equation}

where the induced metric on the visible brane is $g_{v,\mu\nu}=e^{-2kr\pi
}g_{\mu\nu}$, with $g_{\mu\nu}$ the 4d Einstein frame metric. The visible
brane EM action can then be rewritten as%
\begin{equation}
S_{v}=\int d^{4}x\sqrt{-g}\left(  -\frac{1}{4}F^{\mu\nu}F_{\mu\nu}\right)
\label{e71}%
\end{equation}

from which we see an effective dielectric function of unity, $\varepsilon=1$.
We therefore expect no reflection of EM fields from a dimensional boundary in
the visible brane. This qualitative difference in the reflectivity of EM
fields at a dimensional boundary could thus serve to distinguish the RS1
braneworld model from one with a brane-free, warp-free compactified extra
dimension. Similarly, the qualitative and quantitative differences between
various extra dimensional models regarding transmission and reflection of
massless and massive modes could help to differentiate among various models.

\appendix

\section{Numerical Considerations}

Here we discuss some details of the numerical procedure used to extract
solutions for the Klein-Gordon and electromagnetic wave equations (see
previous section), which can be written in terms of a dimensionless frequency
$\bar{\omega}=\omega\delta$ as%
\begin{align}
\phi^{\prime\prime}+\bar{\omega}^{2}\phi-\varepsilon^{-1}\phi &
=0,\label{a1}\\
B^{\prime\prime}+\gamma^{\prime}B^{\prime}+\gamma^{\prime\prime}B+\bar{\omega
}^{2}B  &  =0 \label{a2}%
\end{align}

respectively, where $\varepsilon=\frac{1}{2}\left[  (\varepsilon
_{1}+\varepsilon_{2})+(\varepsilon_{2}-\varepsilon_{1})\tanh\bar{x}\right]  $
and $\gamma=\ln\varepsilon$. The numerical procedure is essentially the same
for either case. We discuss here our procedure for the Klein-Gordon equation.
We assume complex solutions and write $\phi=\phi_{R}+i\phi_{I}$. The real and
imaginary parts of $\phi$ both independently satisfy eq.(\ref{a1}). The
function $\varepsilon$ is constant far away from the transition region
centered at $x=0$. Therefore we assume pure oscillatory solutions in
asymptotic boundary regions far from $x=0$. Let $\phi^{-}$ and $\phi^{+}$ be
the boundary solutions at the positive and negative $x$ boundary regions. The
boundary solutions must then take the form%
\begin{align}
\phi^{-}  &  =e^{ik_{1}x}+A_{1}e^{i\delta_{1}}e^{-ik_{1}x},\label{a3}\\
\phi^{+}  &  =A_{2}e^{i\delta_{2}}e^{ik_{2}x}+A_{3}e^{i\delta_{3}}e^{-ik_{2}%
x}, \label{a4}%
\end{align}

where $k=\sqrt{\bar{\omega}^{2}-\varepsilon^{-1}}$. The boundary condition
$\phi^{-}$ represents our \textquotedblleft initial\textquotedblright%
\ conditions for the numerical integration of the differential equations
associated with $\phi_{R}$ and $\phi_{I}$. The second term in eq.(\ref{a3})
represents the reflected wave. Since we are interested in only right-moving
waves at $x\rightarrow+\infty$, we seek solutions where $A_{3}=0$. Thus we
search for solutions that connect $\phi^{-}$ to $\phi^{+}$ with $A_{3}=0$. A
two parameter search is needed to meet these conditions. The parameter space
$(0<A_{1}<1$,$\ -\pi<\delta_{1}<\pi)$ is searched for the outgoing wave which
has $A_{3}=0$. Numerically, this condition is met in the following manner:
Eq.(\ref{a4}) is examined at $x=\frac{2n\pi}{k_{2}}$ and $x=\frac{(4n+1)\pi
}{2k_{2}}$, where $n$ is a positive integer large enough so that a pure
oscillatory solution is guaranteed. We then get%
\begin{align}
\phi_{R}^{+}(2n\pi/k_{2})  &  =\alpha_{1}=A_{2}\cos\delta_{2}+A_{3}\cos
\delta_{3}\label{a5}\\
\phi_{I}^{+}(2n\pi/k_{2})  &  =\alpha_{2}=A_{2}\sin\delta_{2}+A_{3}\sin
\delta_{3}\label{a6}\\
\phi_{R}^{+}((4n\pi+1)/2k_{2})  &  =\beta_{1}=-A_{2}\sin\delta_{2}+A_{3}%
\sin\delta_{3}\label{a7}\\
\phi_{I}^{+}((4n\pi+1)/2k_{2})  &  =\beta_{2}=A_{2}\cos\delta_{2}-A_{3}%
\cos\delta_{3} \label{a8}%
\end{align}

After a little rearrangement we arrive at the equations we use for the
numerical search:%
\begin{align}
A_{2}  &  =\frac{1}{2}\sqrt{(\alpha_{1}+\beta_{2})^{2}+(\alpha_{2}-\beta
_{1})^{2}}\label{a9}\\
A_{3}  &  =\frac{1}{2}\sqrt{(\alpha_{1}-\beta_{2})^{2}+(\alpha_{2}+\beta
_{1})^{2}} \label{a10}%
\end{align}

Technically, we found that setting $n=5$ and a linear interpolation to the
solutions (\ref{a5})--(\ref{a6}) at the specified $x$ boundary values were of
sufficient numerical accuracy for our purposes. From eq.(\ref{a10}) we see
that the numerical solution must meet the condition that $\alpha_{1}=\beta
_{2}$ and $\alpha_{2}=-\beta_{1}$. The reflection and transmission
coefficients are then given by eq.(\ref{e67}): $\mathcal{R}=A_{1}^{2}$ and
$\mathcal{T}=\frac{k_{2}}{k_{1}}A_{2}^{2}$. Each two parameter search was
conducted at a specific value of $\bar{\omega}$. Numerical integration of the
differential equations for $\phi_{R}$ and $\phi_{I}$ was accomplished using an
Adams Pece integrator in the numerical integration code entitled \textit{DE}
by Shampine and Gordon\cite{DE}. A code was written in Fortran and executed on
IUN's AVIDD-N computer cluster.

\bigskip

\textbf{Acknowledgement:} We are grateful to the Indiana University Computing
Services for the use of the AVIDD-N pentium 3 Linux cluster. All computations
in this paper were conducted on this cluster.

\bigskip

\newpage

\begin{center}
\textbf{Figure Captions}
\end{center}

\bigskip

\textbf{Figure 1:} Reflection and transmission coefficients for
electromagnetic waves. The values for $\varepsilon_{2}$ (2, 5, 10, 20, 100)
are given in the figure. In all cases $\varepsilon_{1}$ was set equal to
1.\textbf{Top:} Reflection coefficient $\mathcal{R}$ as a function of
dimensionless frequency $\bar{\omega}=\omega\delta$. The coefficient agrees
with the theoretical value of $\mathcal{R}(\bar{\omega}=0)=\left(
\dfrac{\varepsilon_{2}-\varepsilon_{1}}{\varepsilon_{1}+\varepsilon_{2}%
}\right)  ^{2}$. \textbf{Bottom:} Transmission coefficient as a function of
$\bar{\omega}$.

\bigskip

\textbf{Figure 2:} Reflection and transmission coefficients for massive
spinless bosons. The values of $\varepsilon_{2}$ (1.1, 1.5, 2.0, 5) are given
in the figure except for $\varepsilon_{2}=2$ which is omitted for visual
clarity. In all cases $\varepsilon_{1}$ was set equal to 1. \textbf{Top:}
Reflection coefficient $\mathcal{R}$ as a function of dimensionless frequency
$\bar{\omega}=\omega\delta$. Numerically we found $\mathcal{R}\rightarrow1$ as
$\bar{\omega}\rightarrow1$. \textbf{Bottom:} Transmission coefficient
$\mathcal{T}$ as a function of $\bar{\omega}$.

\newpage%

%TCIMACRO{\FRAME{dtbpF}{4.8646in}{8.4855in}{0pt}{}{}{fig1a.eps}%
%width 4.8646in;  height 8.4855in;  depth 0pt;  original-width 4.8369in;
%original-height 8.4579in;  cropleft "0";  croptop "1";  cropright "1";
%cropbottom "0";  filename 'fig1a.eps';file-properties "XNPEU";}}}%
%BeginExpansion
\begin{center}
\includegraphics[
height=8.4855in,
width=4.8646in
]%
{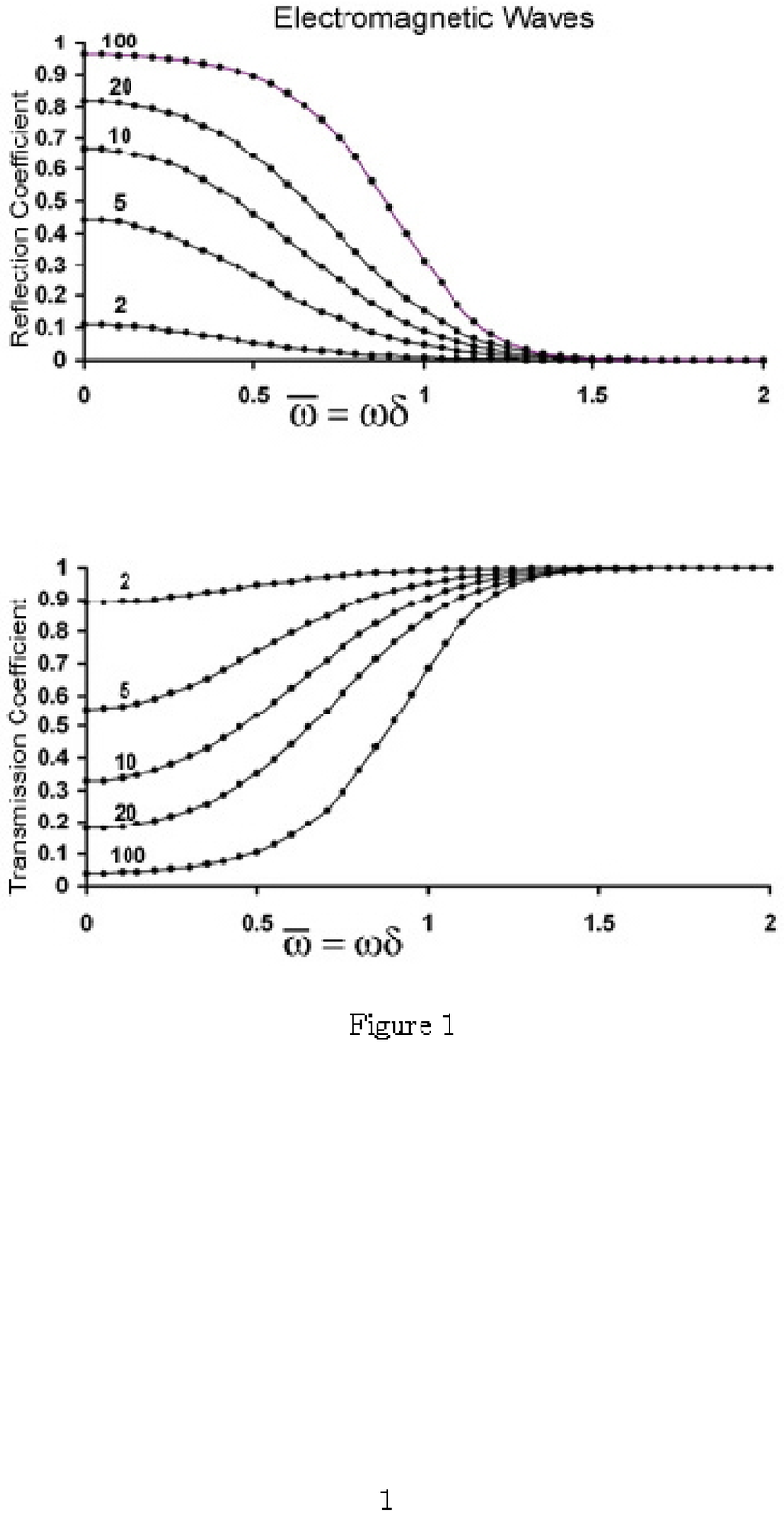}%
\end{center}
%EndExpansion

\newpage%

%TCIMACRO{\FRAME{dtbpF}{4.8646in}{8.4855in}{0pt}{}{}{fig2a.eps}%
%width 4.8646in;  height 8.4855in;  depth 0pt;  original-width 4.8369in;
%original-height 8.4579in;  cropleft "0";  croptop "1";  cropright "1";
%cropbottom "0";  filename 'fig2a.eps';file-properties "XNPEU";}}}%
%BeginExpansion
\begin{center}
\includegraphics[
height=8.4855in,
width=4.8646in
]%
{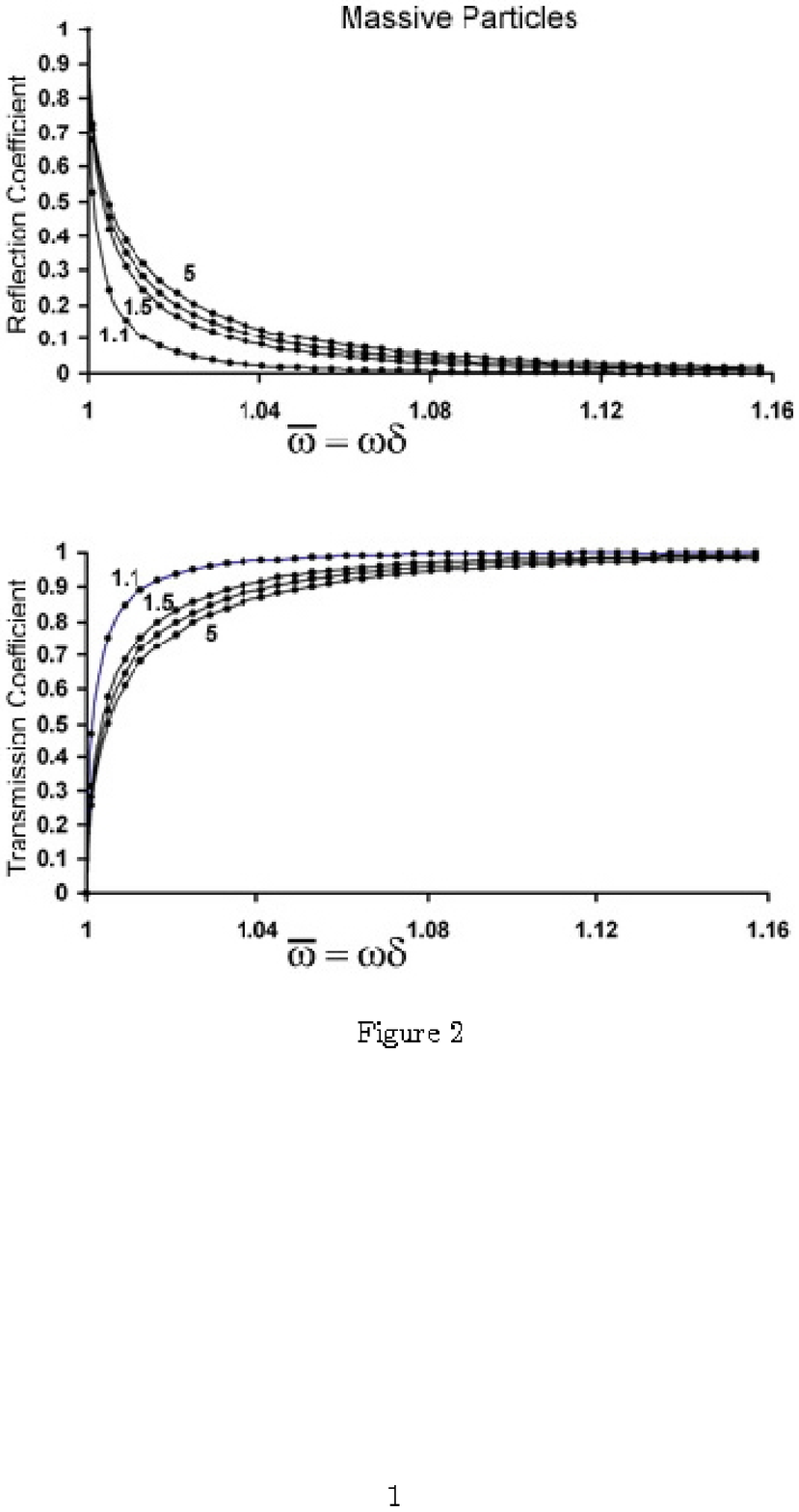}%
\end{center}
%EndExpansion

\end{document}